\documentclass[11pt]{article}
\usepackage{latexsym}
\usepackage{a4}
\usepackage[dvips]{graphicx}
\usepackage{mathrsfs}
\usepackage{color}
\usepackage{bbm}       % double-striked fonts for sets of integer numbers etc.
\newcommand{\MM}{\mathsf{M}}
\newcommand{\WW}{\mathsf{W}}

\newcommand{\no}{\noindent}
\newcommand{\be}{\begin{eqnarray}}
\newcommand{\ee}{\end{eqnarray}}

\newcommand{\up}[1]{\raisebox{1.5ex}[-1.5ex]{#1}}

\begin{document} 
\begin{titlepage}
  \vspace{1.5cm}

%%%%%%%%%%%%%%%%%%%%%%%%%%%%%%%%%%%%%%%%%%%%%%%%%%%%%%%%%%%%%%%%%%%%%%%%%%%%%%%

\renewcommand{\thefootnote}{\fnsymbol{footnote}} 

\begin{center}
\Large\bfseries
Center Vortex Model for the Infrared Sector of $SU(3)$ Yang-Mills Theory
-- Vortex Free Energy
\end{center}
\vspace{1cm}

%%%%

\renewcommand{\thefootnote}{\fnsymbol{footnote}} 
\centerline{M.~Quandt\footnote[1]{Supported by 
             \emph{Deutsche Forschungsgemeinschaft} under contract 
             \# DFG-Re 856/4-2.}, 
             H.~Reinhardt${}^\ast$\vspace{0.5cm}}
\centerline{\textit{Institut f\"ur Theoretische Physik, 
             Universit\"at T\"ubingen} }
\centerline{\textit{D--72076 T\"ubingen, Germany.}\vspace{1cm}} 
\centerline{M.~Engelhardt\footnote[2]{Supported by the U.S.~DOE
under grant number DE-FG03-95ER40965.}\vspace{0.5cm}}
\centerline{\textit{Physics Department, New Mexico State University}}
\centerline{\textit{Las Cruces, NM 88003, USA}}
\renewcommand{\thefootnote}{arabic{footnote}} \vspace{1.5cm}

%%%%%%%%%%%%%%%%%%%%%%%%%%%%%%%%%%%%%%%%%%%%%%%%%%%%%%%%%%%%%%%%%%%%%%%%%%%%%%%%

\begin{abstract}
The vortex free energy is studied in the random vortex world-surface model
of the infrared sector of $SU(3)$ Yang-Mills theory.
The free energy of a center vortex extending into two spatial directions,
which is introduced into Yang-Mills configurations when acting with
the 't~Hooft loop operator, is verified to furnish an order 
parameter for the deconfinement phase transition. It is shown to exhibit 
a weak discontinuity at the critical temperature, corresponding to the weak 
first order character of the transition.
\end{abstract}
\vspace{1cm}

{\footnotesize
}

%%%%

\end{titlepage}

%%%%%%%%%%%%%%%%%%%%%%%%%%%%%%%%%%%%%%%%%%%%%%%%%%%%%%%%%%%%%%%%%%%%%%%%%%%%%%

\section{Introduction} 
\label{sec:1}
\no

Yang-Mills theory in four space-time dimensions undergoes a deconfinement
phase transition at non-zero temperatures. One order parameter used in
practice to characterize
this transition is the \emph{Polyakov loop}, which can be interpreted
as representing a single static quark, immersed in a thermal Yang-Mills
background. Since the confining phase is characterized by the free
energy $F_q$ of such a quark being infinite, the Polyakov loop
$\langle P\rangle = \exp(-F_q)$ is expected to vanish below the critical
temperature $T_c$ , while it should be non-zero in the deconfined
(high temperature) phase. The Polyakov loop is closely related to the
so-called \emph{center symmetry}, which is broken at high temperatures
$T>T_c$. Physically, this critical behavior is counterintuitive; usually,
e.g.~in most spin models, the high temperature phase 
is maximally symmetric and the symmetries are broken \emph{below} the critical
temperature. Furthermore, when formulated on a compact space with periodic
boundary conditions (such as used in lattice calculations), Yang-Mills
theory strictly speaking does not admit a single quark as a physical
excitation, regardless of whether confinement is realized or not.

Not least due to these issues, it is instructive to consider an alternative
dual (dis-)order parameter to characterize the deconfinement phase transition,
namely the free energy associated with a center vortex world-surface.
Center vortices are defined by the property that they contribute a center
phase to a Wilson loop when piercing an area spanned by the latter.
The center vortex free energy is expected to behave in a manner which
is dual to the behavior of Wilson loops:
\begin{itemize}
\item The free energy of vortex world-surfaces extending into one space
direction and the (Euclidean) time direction is expected to exhibit no
area-law dependence on the vortex world-surface area at all temperatures
$T$, if the corresponding dual (spatial) Wilson loops exhibit an area law for
all $T$ (this is the case for pure $SU(N)$ Yang-Mills theory).
\item The free energy of vortex world-surfaces extending into two spatial
directions is expected to behave in a manner dual to the behavior of
corresponding (temporal) Wilson loops: No area-law dependence of the free
energy on the vortex world-surface area at temperatures below the
deconfinement transition temperature $T_c$, but behavior according to
an area law above the critical temperature.
\end{itemize}
The latter observation permits the definition of a (temperature-dependent)
\emph{dual string tension} $\tilde{\sigma}$ through the leading behavior
of the excess free energy $F_P $ in the presence of asymptotically large
vortex world-surfaces $P$,
\begin{equation}
F_P \equiv -\ln (Z_P /Z_0 ) \,\,\longrightarrow\,\, \tilde{\sigma}(T)\,
\mathrm{Area}(P) \,,\qquad T > T_c
\label{eq1}
\end{equation}
which can serve as an order parameter for the deconfinement phase transition.
Here, $Z_0 $ denotes the conventional Yang-Mills partition function, 
whereas $Z_P $ denotes the partition function in the presence of the 
additional vortex $P$.

Using the vortex free energy as a confinement order parameter was first
suggested by `t~Hooft \cite{R3}, who, initially working in the
Schr\"odinger picture (using the Weyl gauge), defined a magnetic loop
{\em operator} $\MM(\mathscr{C})$ in the continuum
\cite{R3} via equal-time commutation relations with \emph{all} spatial
Wilson loops $\WW(\mathscr{C}')$,
\begin{equation}
\MM(\mathscr{C})\,\WW(\mathscr{C}')\,\MM(\mathscr{C})^\dagger = z(\mathscr{C},
\mathscr{C}') \,\WW(\mathscr{C}')\,.
\label{1}
\end{equation}
Here, $z(\mathscr{C},\mathscr{C}')$ is an element of the center of the
gauge group related to the integral \emph{linking number}
$\nu(\mathscr{C},\mathscr{C}')$ of the two loops
$\mathscr{C}$ and $\mathscr{C'}$ in three space dimensions,
\begin{equation}
z(\mathscr{C},\mathscr{C}') = \exp\left(i \frac{2\pi}{N}\,
\nu(\mathscr{C},\mathscr{C}')\right)\qquad\qquad
( \, SU(N)\, \mbox{gauge group} \, ).
\label{2}
\end{equation}
The loop operator $\MM$ of course is only defined
\emph{implicitly} by the commutation relation eq.~(\ref{1}).
An \emph{explicit} realization of this formal construction
was recently derived and
discussed in detail in ref.~\cite{R7}, cf.~also \cite{kortplb}. Also in the
dual language, the change of behavior of the vortex free energy across
the deconfinement phase transition can be associated with a (magnetic)
center symmetry \cite{R3,kortsym}, which is broken in the
{\em low-temperature} phase.

Subsequent to `t~Hooft's suggestion, realizations of the `t~Hooft loop
operator within the framework of lattice gauge theory formulated in
3+1-dimensional (Euclidean) space-time were discussed and schemes of
measuring the free energies of the vortices thus introduced into the gauge
configurations were devised \cite{mack,uwg,tomold,debbio1}. On this basis,
sufficient conditions for confinement via vortex condensation were established
\cite{mack,tomineq}. While the construction given in \cite{mack} admits the
introduction of vortices of a large variety of shapes into gauge
configurations (by imposing suitable boundary conditions on ``vortex
containers'' within the lattice), in practice a particularly
clean way to investigate the center vortex free energy is
to consider specifically a vortex world-surface which occupies a
complete two-dimensional plane in space-time\footnote{Note that the
covariant setting given by the lattice gauge theory framework permits
the definition of vortex world-surfaces extending in arbitrary directions
in space-time; however, ones extending purely in two spatial directions
are of course the relevant ones as far as defining an order parameter
for the deconfinement phase transition is concerned.}. On a compact
space-time, such as used in lattice Yang-Mills theory, such a world-surface
is closed by virtue of the periodic boundary conditions; furthermore,
its introduction corresponds to adopting twisted boundary conditions in the
directions orthogonal to the vortex \cite{thoo2}.
This is reflected in the fact that configurations with such a vortex
are not connected to the conventional vacuum by the dynamics. The dynamics
of Yang-Mills theory and also of the random vortex world-surface model
considered in this work, cf.~\cite{R1,R2}, can create at most pairs of
vortices occupying complete two-dimensional (parallel) planes in space-time;
a single such vortex is topologically stable. Also, its free energy is not
contaminated by ancillary effects such as the ones which occur, e.g., when
one considers open vortex world-surfaces; these are bounded by center
monopole world-lines, and their free energies thus contain additional
contributions stemming from monopole self-energies and interactions, which
need to be disentangled from the vortex free energy itself\footnote{A note
is in order concerning the ``center monopole'' concept. Center monopoles
in the strict sense, i.e., sources and sinks of magnetic flux of magnitude
$2\pi /N$, are unphysical objects, since they violate the Bianchi identity
(continuity of magnetic flux modulo $2\pi $). In lattice Yang-Mills theory,
when separating the gauge fields into center and coset components, one can
identify locations at which magnetic flux carried by the center component
spreads out and continues as part of the coset component \cite{tomold}.
These locations are often called ``center monopoles'', although strictly
speaking they are not sources or sinks of magnetic flux and do not violate
the Bianchi identity. On the other hand, in the vortex model discussed here,
there is no analogue of the coset; center vortices (albeit implicitly
endowed with a finite thickness) are the only degrees of freedom, and
open vortex world-surfaces, bounded by center monopoles, are unphysical.
Nevertheless, further below, open vortex surface configurations will be
used as intermediate objects in setting up the numerical calculation; these
should not be viewed as any more than auxiliary mathematical constructs.
The dual string tension measured will ultimately be derived from the free
energy of a closed vortex world-surface.}.

A number of measurements of vortex free energies have been carried out
within lattice Yang-Mills theory, in a variety of guises. The excess free
energy in the presence of twisted boundary conditions compared to
periodic boundary conditions was evaluated in
\cite{mpi,kaj1,tomb,hart,forcpep,lor1,lor2,forcjahn,vetto}.
Also the free energy of more varied vortex configurations, including open
vortex world-surfaces \cite{debbio1,forcpep,rebbi,debbio2}
and intersecting planar vortices \cite{lor1} was investigated. Partition
functions in the presence of twists can furthermore be related to electric
flux free energies via $Z(N)$ Fourier transforms \cite{thoo2}; a thorough
discussion of this together with detailed measurements can be found in
\cite{lor1}, cf.~also \cite{hasen,gonmart} for earlier work with this
focus. In addition, in the case of a system with a first order deconfinement
transition, as found, e.g., for $SU(3)$ color, the vortex free energy in the
deconfined phase is related to the $Z(3)$ order-order interface tension
\cite{kortplb}; for an evaluation of this interface tension using a different
method, cf.~\cite{huang2}. With auxiliary information about wetting
properties \cite{huang2,gross2}, the order-order interface tension can in
turn be connected to the order-disorder interface tension at the critical
temperature. Direct measurements of the order-disorder interface tension
at the $SU(3)$ deconfinement transition were reported in
\cite{gross2,kaj2,huang1,gross1,iwasaki,aoki,beinlich,papa};
for a recent comparison with data obtained using a new calculational method
based on the vortex free energy, cf.~\cite{vetto}. The results obtained in
the present work will be put in relation to the survey presented in
\cite{vetto} in section \ref{discuss}.

\section{Vortex free energy in the random vortex world-surface model}
The main purpose of the work presented here is to compute the dual string
tension $\tilde{\sigma}(T)$ by evaluating the free energy of appropriate
vortex world-surfaces within the $SU(3)$
random vortex world-surface model. This model was first defined and studied
for the gauge group $SU(2)$ in refs.~\cite{R1,preptop} and later
extended to the $SU(3)$ case in refs.~\cite{R2,bary}. It describes the
infrared sector of Yang-Mills theory as an ensemble of random (thick)
center vortex world-surfaces. On a space-time lattice dual to the one
Yang-Mills lattice links are defined on\footnote{Two lattices of spacing $a$
are dual to one another if one can be generated by shifting the other one
by the vector $(1,1,1,1)a/2$.}, the vortex world-surfaces are composed of
elementary squares, and can thus link with the Wilson loops
of the original lattice. As described and motivated in detail in
\cite{R1,R2}, since the model is not intended to describe the structure
of the theory at arbitarily short distances, it has a fixed lattice
spacing $a$ representing the vortex thickness.
Continuity of the magnetic flux forces the vortex world-surfaces to be 
closed, and the statistical weight of a (closed) vortex world-surface is 
determined by a model action inspired by a gradient expansion of the 
Yang-Mills action of a center vortex \cite{Rx}:
\begin{eqnarray}
S &=& \epsilon \sum_x\sum_{\mu,\nu \atop \mu<\nu} |q_{\mu\nu}(x)| + 
c \sum_x\sum_\mu \Bigg[ \sum_{\nu < \lambda \atop \nu \neq \mu,
\lambda\neq \mu} \Big( | q_{\mu\nu}(x) \, q_{\mu\lambda}(x) |
 + | q_{\mu\nu}(x) \, q_{\mu\lambda}(x-e_\lambda) | + \nonumber \\
& & \qquad\qquad\qquad\qquad
+ | q_{\mu\nu}(x-e_\nu) \, q_{\mu\lambda}(x) |
 + | q_{\mu\nu}(x-e_\nu) \, q_{\mu\lambda}(x-e_\lambda) |
\Big)\Bigg]  
\label{action}
\end{eqnarray}
Here, the (dual) lattice elementary square extending from the dual lattice
site $x$ into the positive $\mu $ and $\nu $ directions is associated with a
\emph{triality} $q_{\mu\nu} (x) \in \{-1,0,1\}$ (in the case of an underlying
$SU(3)$ gauge group), labeling the center flux carried by that elementary
square. The value $q_{\mu\nu} (x) =0$ indicates that the square is not part
of a vortex world-surface, whereas the other two values are associated with
the two types of center flux a vortex surface can carry\footnote{To be
precise, in terms of elementary Wilson loops (plaquettes) on the original
lattice, the plaquette $U_{\alpha\beta}(y)$ extending from $y$ into the
positive $\alpha $ and $\beta $ directions takes the value
$U_{\alpha\beta}(y) = 
\exp(i \pi/3 \cdot \epsilon_{\alpha\beta\mu\nu}\,q_{\mu\nu}(x))$, where
$x=y+(e_{\alpha } +e_{\beta } -e_{\mu } -e_{\nu } ) a/2$,
with $e_{\lambda } $ denoting the unit vector in the $\lambda $-direction.}.
The two terms in eq.~(\ref{action}) correspond to a Nambu-Goto and
a curvature term, respectively. They implement the notion that vortices,
on the one hand, may be associated with a surface tension such that it
costs a certain action increment $\epsilon$ to add an elementary square to
the surface. On the other hand, vortices are \emph{stiff}, such
that an action increment $c$ is incurred for each pair of elementary 
squares in the vortex surface which share a link but do not lie in the 
same plane (i.e., vortex curvature is penalised). For details on the 
physical foundations of the model, and the determination of the couplings 
$\epsilon$ and $c$, the reader is referred to refs.~\cite{R1,R2}. In
practice, physical ensembles which reproduce the gross features
of the corresponding Yang-Mills theory in the infrared can be achieved
with $\epsilon =0$.
In the case of an underlying $SU(2)$ gauge group, the value $c=0.24$
generates a physical ensemble, whereas in the case of the $SU(3)$ gauge
group, the appropriate value is $c=0.21$. These values were determined
by requiring the models to correctly reproduce the ratio of the
deconfinement temperature $T_c $ to the square root of the zero-temperature
string tension $\sqrt{\sigma }$ found in the corresponding Yang-Mills
theory.

Since the random vortex world-surface model was proposed as an effective
description of the long-range properties of Yang-Mills theory, including
its deconfinement phase transition, the temperature dependence
of $\tilde{\sigma}(T)$ near the transition is a non-trivial and important
test of the model. In the random vortex world-surface model, the elementary
squares $q_{\mu\nu} (x)$ of the dual lattice represent the building blocks
for the dynamical center vortex degrees of freedom; in this framework,
it is thus particularly simple to introduce an additional vortex
world-surface into the configurations. One simply needs to
replace\footnote{The modulo operation is to be applied such that the
result again takes a value in $\{ -1,0,1 \} $.}
\begin{equation}
q_{\mu\nu} (x) \longrightarrow (q_{\mu\nu} (x) + \bar{q} ) \, \mbox{mod} \, 3
\label{elemtrafo}
\end{equation}
for all elementary squares $q_{\mu\nu} (x)$ making up the world-surface
in question, with a fixed value of $\bar{q} \in \{ -1,1\} $. The two
possible values of $\bar{q} $ are related by a space-time inversion which
reverses the direction of magnetic flux (and leaves the action invariant);
one can therefore restrict oneself to $\bar{q} =1$ without loss of generality.

The dual string tension is obtained in this work specifically by calculating
the excess free energy in the presence of a vortex world-surface $P$
occupying an entire lattice plane extending into two spatial directions.
The plane $P$ is composed of elementary squares $p_i $ with $1\leq i\leq I$,
where $I$ is the number of elementary squares in $P$. Let $Q$ denote
collectively a vortex world-surface configuration $\{ q_{\mu\nu} (x)\} $,
and $S[Q]$ the corresponding action (\ref{action}). It will be useful for
the following to introduce a notation for introducing additional elementary
vortex squares $p_i $ into configurations; this is best done recursively:
Starting with a configuration $Q_0 $ from the conventional random vortex
world-surface ensemble, without any additional vortex surfaces introduced,
define the configuration $Q_i $ as the configuration obtained by 
effecting the transformation (\ref{elemtrafo}) specifically on the
elementary square $p_i $ of the configuration $Q_{i-1} $. One can also
define corresponding partition functions $Z_i $ as
\begin{equation}
Z_i = \int [dQ_i ] \exp (-S[Q_i ] ) = \int [dQ_0 ] \exp (-S[Q_i ] ) \ .
\end{equation}
In the notation of eq.~(\ref{eq1}), $Z_P \equiv Z_I $. To obtain the dual
string tension, it is thus necessary to evaluate
\begin{equation}
\frac{Z_I }{Z_0 } = \frac{\int [dQ_0 ] \exp (-S[Q_0 ] )
\exp (-(S[Q_I ] - S[Q_0 ]) )}{\int [dQ_0 ] \exp (-S[Q_0 ] )}
\equiv \left\langle \exp (-(S[Q_I ] - S[Q_0 ]) ) \right\rangle_{0} \ .
\end{equation}
However, this expression is of little practical use as it suffers from a
serious \emph{overlap problem}: The quantity being averaged varies over
many orders of magnitude, so that most configurations only give an
\emph{exponentially small} contribution to the average. The resulting
numerical noise precludes extracting a useful signal. This problem is
addressed by using an algorithm introduced by de~Forcrand et al
\cite{forcpep}. By decomposing
\begin{equation}
\frac{Z_I }{Z_0 } = \frac{Z_I}{Z_{I-1}}\cdot
\frac{Z_{I-1}}{Z_{I-2}} \cdots\frac{Z_1}{Z_0} \ ,
\label{decomp}
\end{equation}
the problem is separated into the calculation of a (sizeable) number of
independent expectation values with good overlap; namely, one evaluates the
effect of introducing just one additional elementary square $p_i $ into the
configurations at a time,
\begin{equation}
\frac{Z_i}{Z_{i-1}} = \frac{\int [dQ_0 ] \exp (-S[Q_{i-1} ] )
\exp (-(S[Q_i ] - S[Q_{i-1} ]) )}{\int [dQ_0 ] \exp (-S[Q_{i-1} ] )}
\equiv \left\langle \exp (-(S[Q_i ] - S[Q_{i-1} ]) ) \right\rangle_{i-1} \ .
\label{indiv}
\end{equation}
Note that the classes of configurations $Q_i $, excepting $Q_0 $ and $Q_I$,
contain open vortex world-surfaces, bounded by center monopoles, which
violate the Bianchi identity. As already discussed further above, these
are unphysical objects which are introduced here merely as intermediate
mathematical constructs. If one implemented the Bianchi constraint in terms
of an infinite action term penalizing center monopoles, all intermediate
partition functions $Z_i $ (apart from $Z_0 $ and $Z_I$) strictly speaking
would vanish. These $Z_i $ thus really are calculated with a modified weight
in which the infinite self-energies of the center monopoles introduced
explicitly into the configurations are left out (all other center monopoles
are of course still completely suppressed; the integration $\int [dQ_0 ] $
is over the conventional ensemble which respects the Bianchi identity).
Of course, the final result for the dual string tension ultimately depends
only on the physical partition functions $Z_I $ and $Z_0 $; all intermediate
partition functions cancel out in the product (\ref{decomp}).

In practice, it is possible to improve the measurement of the individual
expectation values (\ref{indiv}) further by using a multi-hit procedure.
The quantity $S[Q_i ] - S[Q_{i-1} ]$ depends only on the configuration
in the neighborhood of the elementary square $p_i $; thus, by performing
multiple configuration updates and measurements in that neighborhood before
carrying out the next global update of the configuration, statistics can
be improved considerably without spoiling detailed balance or ergodicity.

As the plane $P$ is filled up one elementary square $p_i $ at a
time\footnote{In practice, this was done row by row, as one would read a
text.}, one can record the free energies
\begin{equation}
F_i = -\ln (Z_i /Z_0 )
\label{eq11}
\end{equation}
of the partial vortex surfaces, up to the free energy $F_I $ of the full
plane $P$. The dual string tension is extracted as
\begin{equation}
\tilde{\sigma}(T) = \frac{F_I }{Ia^2 }
\end{equation}
where $a$ denotes the lattice spacing. As already mentioned above, the
lattice spacing in the random vortex world-surface model is a fixed
physical quantity related to the transverse thickness of the vortices;
for a detailed discussion, cf.~\cite{R1,R2}. At the physical point
$\epsilon =0, c=0.21$ of the $SU(3)$ model, its value is
$a=0.39\,\mbox{fm} $, where the scale was fixed by equating the
zero-temperature string tension with $(440\,\mbox{MeV})^2 $.
Besides the dual string tension, which is the principal quantity of
interest in this work, below also the partial free energies $F_i $ will be
discussed.

\section{Numerical results}
\label{sec:5}
\subsection{General remarks}
\label{sec:5.1}
Since the lattice spacing $a$ of the random vortex world-surface model
is taken to be a \emph{fixed} quantity that is not scaled towards a 
continuum limit, the temperature in the model can only be changed in rather
large steps if the coupling constants are kept fixed. However, by employing
an interpolation method, a continuous range of temperatures can be explored
\cite{R1,R2}: For all measurements in this work, $\epsilon=0$ is fixed
and only variations in the curvature coupling $c$ are considered. For
several temporal extensions $N_0 = 1,2,3$ of the lattice, one can vary the
coupling $c$ until the deconfinement phase transition is observed at
critical couplings $c_i^\ast \equiv c^\ast(N_0 = i)$. The critical couplings
for the gauge group $SU(3)$ on lattices of extension $N_0 \times 30^3$
are\footnote{The value of $c^\ast_3 $ was only determined to within an error
of 1\% due to its weak influence on the interpolations needed in practice.
The other two values are accurate to the digits shown.}

\medskip
\[
\begin{tabular}{|c||c|c|c|}
\hline
$N_0 = i$  & 1 & 2 & 3 \\\hline\hline
$c^\ast_i$ & 0.0872 & 0.2359  & 0.335  \\\hline
\end{tabular}
\]

\bigskip
From the critical temperature $a T_c = 1 / N_0$ for the three values 
$c=c_i^\ast$, the function $a\,T_c(c)$ can be determined for all relevant
couplings $c$ by interpolation. This, in turn, specifies the temperature
in units of the deconfinement transition temperature for any $c$ and $N_0$
via
\[
\frac{T}{T_c} = \frac{1}{N_0\cdot a T_c(c)}\,.
\]
In particular, this permits carrying out measurements at a given $T/T_c $
for different $N_0$ and the corresponding $c$. Combining these 
measurements, the quantity in question, at the given $T/T_c $, can then also 
be obtained for the physical value $c=0.21$ by interpolation in $c$. This 
procedure is used to arrive at the last two columns of Table~\ref{tab2} below.

To be consistent with this determination of the temperature, and also
in order to minimize finite-size effects, the evaluation of the dual
string tension in this work was likewise carried out on $N_0 \times 30^3$
lattices. To check for finite-size effects, control measurements were
performed on lattices with spatial extensions of $10$, $12$ and $16$
lattice spacings; in all cases, the discrepancies between measurements
of the dual string tension at spatial extensions $16$ and $30$, respectively,
were smaller than the statistical errors of the measurements.

\subsection{Boundary effects on open vortex world-surfaces}
\label{sec:5:2}

\begin{figure}[t]
\centerline{
\begin{minipage}{7.5cm}
\includegraphics[width = 7.5cm]{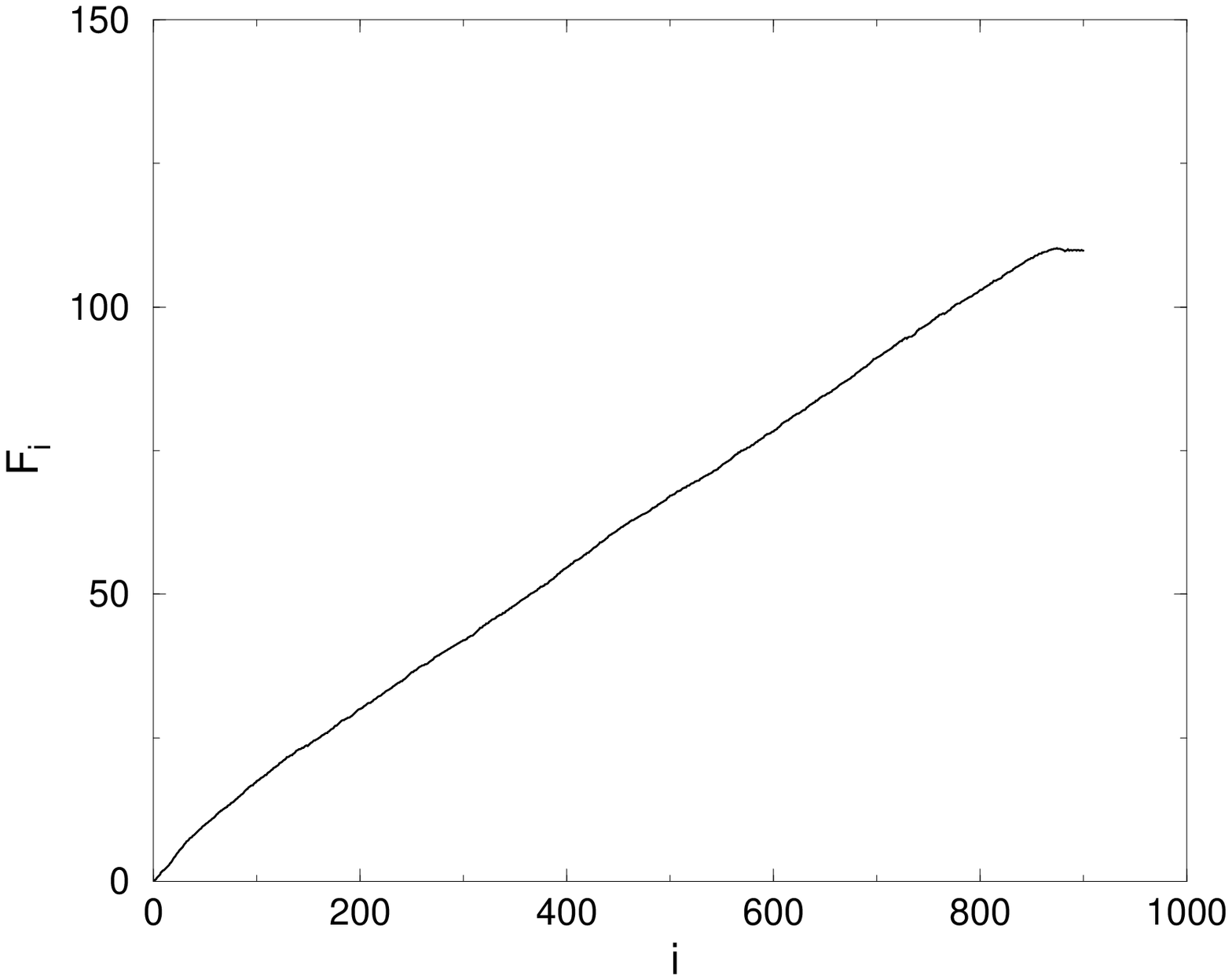} \end{minipage}\hfill 
\begin{minipage}{7.5cm}
\includegraphics[width = 7.5cm]{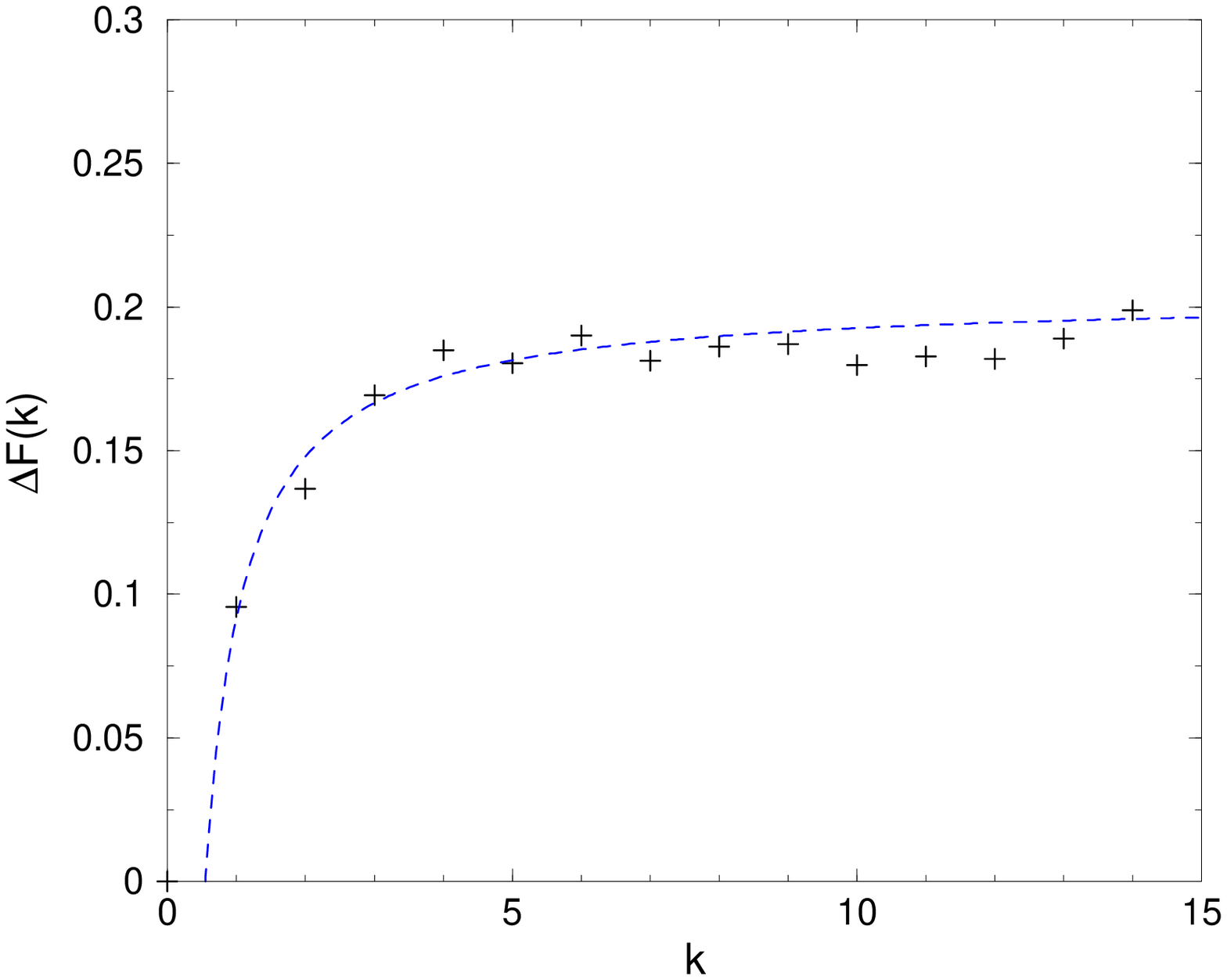}\end{minipage}
}
\caption{Left panel: Free energies $F_i $, cf.~eq.~(\ref{eq11}), of
partial vortex world-surfaces, as a function of the number $i$ of elementary
squares included in the surface. Measurements were taken on a
$30^3 \times 2$ lattice at $\epsilon=0$ and $c=0.2550$, which corresponds
to $T/T_c$ = 1.093 (deconfined phase).
Right panel: The residual magnetic monopole correlation
$\Delta F (k) = F_{kN_s } - F_I \cdot kN_s /I$
as a function of the monopole pair separation in lattice units $k$,
from the same measurement. The curve represents a fit by a (self-energy)
constant plus a Yukawa potential, $\Delta F (k) = V_0 +\exp (-\mu k)/k$.}
\label{fig4}
\end{figure}

The left-hand panel of Fig.~\ref{fig4} shows a typical result for the 
free energies $F_i $  of partial vortex world-surfaces in the deconfined
phase, at $T/T_c = 1.093$. The area of the surfaces in units of $a^2$
is $i$, and the approximate linear rise with $i$ thus indicates an
area law, as expected in the deconfined phase. However, there are also 
systematic deviations from a behavior purely proportional to the area,
in particular for small $i$, and for large $i$, when the surface almost 
completely covers the lattice plane it is located in.

These deviations can be formally attributed to self-energies and residual
interactions of the center monopole world-lines which necessarily bound
the intermediate open vortex world-surfaces. They are the most
transparent when considering surfaces made up of full rows of elementary
squares, i.e., when $i$ is an integer multiple of the spatial extension
of the lattice, $i=kN_s $. In this case, there are two parallel center
monopole world-lines winding along a spatial dimension of the lattice
at a distance $k\cdot a$ and the deviations from the area law formally
correspond to self-energies and residual interactions of the two monopoles.
The right-hand panel in Fig.~\ref{fig4} displays the deviation
$\Delta F (k) = F_{kN_s } - F_I \cdot kN_s /I$ from an area law as a
function of the center monopole separation in lattice units $k$. It
should again be noted that this quantity does not have an immediate
physical interpretation. Apart from the monopoles propagating into a
spatial instead of the temporal direction, they strictly speaking are
associated with an infinite self-energy preventing their appearance
in the physical vortex ensemble. An additional finite contribution to
this infinite self-energy is physically inconsequential.

\subsection{The dual string tension}
Approaching the deconfinement phase transition temperature $T_c $ from
below, one can verify that the dual string tension is compatible with
zero. Corresponding measurements on $30^3 \times 2$ lattices at couplings
$c$ below the critical coupling $c=0.2359$, which realize the confining
phase, yield:

\medskip
\[
\begin{tabular}{|c|c|}
\hline
$c$ & $\tilde{\sigma}\,a^2 $ \\\hline\hline
0.21 & $0.000190 \pm 0.000214$ \\\hline
0.23 & $3.6 \cdot 10^{-6} \pm 0.001$ \\\hline
0.235 & $-0.00037 \pm 0.001$ \\\hline
\end{tabular}
\]

\bigskip
Also, for comparison, a measurement on a $16^3 \times 2$ lattice at $c=0.21$
resulted in $\tilde{\sigma}\,a^2 = -0.000205 \pm 0.000271 $.

\begin{figure}[t]
\centerline{
\includegraphics[width = 10cm]{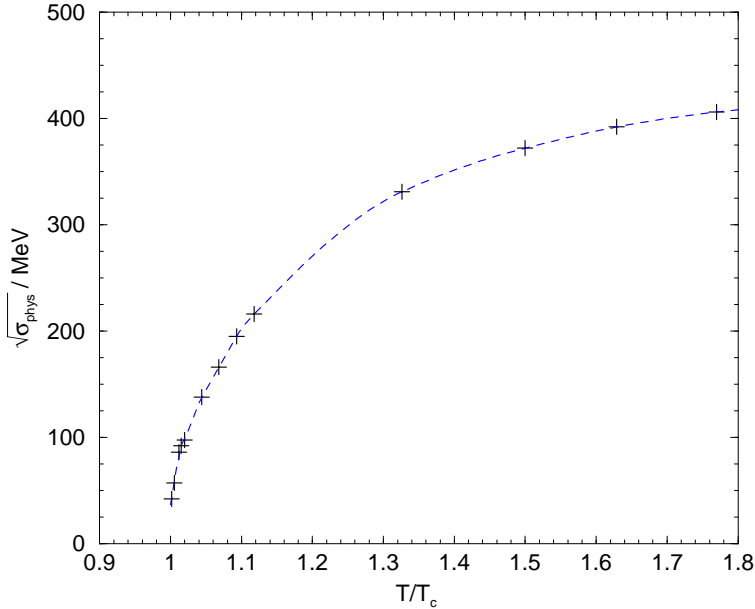} 
}
\caption{Dual string tension $\tilde{\sigma}$ as a function of the
temperature $T$ in the deconfined phase, taken from Table~\ref{tab2}.}
\label{fig6}
\end{figure}

Above the phase transition temperature, measurements were carried out
for a sequence of temperatures $T/T_c $. As displayed in Table~\ref{tab2},
for each temperature, measurements were performed both on a $30^3 \times 2$
and a $30^3 \times 1$ lattice with the appropriate couplings $c$ realizing
the given temperature. Thus, for a fixed temperature $T/T_c$, the dual
string tension $\tilde{\sigma}$ for the two values $c_1$ ($N_0 = 1$) and
$c_2$ ($N_0 = 2$) is obtained. To arrive at the physical value, those
two results are then interpolated linearly to the physical coupling 
$c_{\rm phys} = 0.21$. Since the lattice spacing at the physical point
is known in absolute units, the physical value of the dual string tension
can also be given in absolute units. This is shown in the last column of
Table~\ref{tab2}. These results are also presented graphically in
Fig.~\ref{fig6}. 

\begin{table}
\medskip
\centerline{
\begin{tabular}{|c||c|c|c|c||c|}\hline
$T/T_c$ &  & $c$  & $\tilde{\sigma}\,(30a)^2$ & 
$\tilde{\sigma}_{\rm phys}\,(30a)^2$ & 
$\sqrt{\tilde{\sigma}_{\rm phys}}$ / MeV \\ 
\hline\hline
& $N_0 = 1$ & 0.08753 & 4.9  &	      & 	   \\ \cline{2-4}  
\up{1.0015} & $N_0 = 2$ & 0.2362  & 6.6  & \up{6.3}   & \up{42.3}  \\ \hline 
& $N_0 = 1$ & 0.0884  & 14.4 &	      & 	   \\ \cline{2-4}  
\up{1.0052} & $N_0 = 2$ & 0.2370  & 10.8 & \up{11.5}  & \up{57.1}  \\ \hline 
& $N_0 = 1$ & 0.0899  & 29.3 &	      & 	   \\ \cline{2-4}  
\up{1.0118} & $N_0 = 2$ & 0.2384  & 25.5 & \up{26.2}  & \up{86.2}  \\ \hline 
& $N_0 = 1$ & 0.0905  & 35.6 &	      & 	   \\ \cline{2-4}  
\up{1.0147} & $N_0 = 2$ & 0.2390  & 28.7 & \up{30.0}  & \up{92.2}  \\ \hline 
& $N_0 = 1$ & 0.0916  & 46.7 &	      & 	   \\ \cline{2-4}  
\up{1.0195} & $N_0 = 2$ & 0.2400  & 30.0 & \up{33.4}  & \up{97.3}  \\ \hline 
& $N_0 = 1$ & 0.0969  & 98.1 &	      & 	   \\ \cline{2-4}  
\up{1.044}  & $N_0 = 2$ & 0.2450  & 57.3 & \up{66.9}  & \up{138}   \\ \hline 
& $N_0 = 1$ & 0.1021  & 144  &	      & 	   \\ \cline{2-4}  
\up{1.068}  & $N_0 = 2$ & 0.2500  & 80.2 & \up{97.5}  & \up{166}   \\ \hline 
& $N_0 = 1$ & 0.1073  & 188  &	      & 	   \\ \cline{2-4}  
\up{1.093}  & $N_0 = 2$ & 0.2550  & 110  & \up{134}   & \up{195}   \\ \hline 
& $N_0 = 1$ & 0.1123  & 226  &	      & 	   \\ \cline{2-4}  
\up{1.118}  & $N_0 = 2$ & 0.2600  & 133  & \up{164}   & \up{216}   \\ \hline 
& $N_0 = 1$ & 0.1495  & 437  &	      & 	   \\ \cline{2-4} 
\up{1.326}  & $N_0 = 2$ & 0.3000  & 308  & \up{385}   & \up{330}   \\ \hline 
& $N_0 = 1$ & 0.1756  & 513  &	      & 	   \\ \cline{2-4}  
\up{1.500}  & $N_0 = 2$ & 0.3350  & 387  & \up{486}   & \up{371}   \\ \hline 
& $N_0 = 1$ & 0.1929  & 548  &	      & 	   \\ \cline{2-4}  
\up{1.629}  & $N_0 = 2$ & 0.4500  & 435  & \up{540}   & \up{391}   \\ \hline 
& $N_0 = 1$ & 0.2100  & 579  &	      & 	   \\ \cline{2-4}  
\up{1.770} & $N_0 = 2$  &  ---    & ---  & \up{579}   & \up{405}   \\ \hline 
\end{tabular}}
\caption{\small Dual string tension $\tilde{\sigma}$ measured on
$30^3 \times N_0$ lattices with $\epsilon = 0$. The curvature couplings
$c$ were chosen such that the corresponding measurements for $N_0=1$ and
$N_0=2$ implement the same temperature $T/T_c$. The results in the last
two columns represent the interpolation to the physical value $c=0.21$.
Since the lattice spacing at this point is fixed to $a=0.39\,\mathrm{fm}$,
the value of the physical dual string tension $\tilde{\sigma}_{\rm phys}$
can be given in absolute numbers.}
\label{tab2}
\end{table}

As can be seen from the data, the dual string tension $\tilde{\sigma}$ 
rises with increasing temperature in the deconfined phase. As the phase
transition is approached from above, the dual string tension
$\tilde{\sigma}$ quickly vanishes as is expected for an order parameter.
Since the phase transition in the $SU(3)$ case is weakly first
order \cite{R2}, a \emph{weak discontinuity} is expected in
the dual string tension $\tilde{\sigma}(T)$. The discontinuity can be
obtained by using the two data points nearest to $T_c$ and extrapolating
linearly to $T=T_c$. The statistical uncertainty
for the dimensionless combination $\tilde{\sigma}_{\rm phys} \,(30 a)^2$
(the second to last column in Tab.~\ref{tab2}) is 
$6.3 \pm 0.4$ at $T/T_c = 1.0015$  and 
$11.5 \pm 0.5$ at $T/T_c = 1.0052$. Varying the slope of the extrapolation 
line within these error bounds, the resulting value at $T=T_c+$ is
\begin{equation}
\left.\sqrt{\tilde{\sigma}_{\rm phys}} \right|_{T=T_c+} = 
(34.5 \pm 3.3)\,\mathrm{MeV}\,.
\label{88}
\end{equation}
As a cross-check, a spline extrapolation through the five data points
nearest to the transition (up to $T/T_c = 1.0195$) yields
$\sqrt{\tilde{\sigma}_{\rm phys}} |_{T=T_c+} \approx 36.8\,\mathrm{MeV} $,
whereas using only the three nearest data points leads to
$\sqrt{\tilde{\sigma}_{\rm phys}} |_{T=T_c+} \approx 36.5\,\mathrm{MeV} $,
both consistent with (\ref{88}). Also, linear regression through the nearest
three data points yields 
$\sqrt{\tilde{\sigma}_{\rm phys}} |_{T=T_c+} \approx 35.5\,\mathrm{MeV} $.
The data thus provide a clear signal for the discontinuity of the dual
string tension $\tilde{\sigma} $ at $T=T_c $, in accordance with the first
order character of the phase transition previously verified for the
$SU(3)$ random vortex world-surface model via the action density
distribution at the critical temperature \cite{R2}. The small value
of the discontinuity compared with typical strong interaction scales
again demonstrates the weak first order character of the transition.

\section{Discussion}
\label{discuss}
The dual string tension in the framework of the $SU(3)$ random vortex
world-surface model was verified in this investigation to represent
an order parameter for the deconfinement phase transition, and to
furthermore reflect the weak first order character of the transition.
Quantitatively, the result (\ref{88}) for the discontinuity at the
transition,
\[
\left.\sqrt{\tilde{\sigma}} \right|_{T=T_c+} =
(34.5 \pm 3.3)\,\mathrm{MeV}\,,
\]
is smaller than one would expect from measurements of the order-disorder
interface tension in $SU(3)$ lattice Yang-Mills theory assuming perfect
wetting. The available data for the order-disorder tension \cite{vetto}
indicate a value of $\sigma_{od} \approx 0.015\, T_c^2 $, which translates
to $\sigma_{od} \approx (34\, \mbox{MeV})^{2} $, if the scale is set by
equating the zero-temperature string tension with $(440 \, \mbox{MeV})^{2} $.
Assuming perfect wetting, the order-order interface tension, equivalent
to the dual string tension, would be expected to take the value
$2\sigma_{od} \approx (48\, \mbox{MeV})^{2} $.

There are, however, caveats to this comparison. On the one hand, though
wetting has been observed in $SU(3)$ lattice Yang-Mills theory
\cite{huang2,gross2}, there remains an uncertainty as to its quantitative
extent. On the other hand, there are also still significant uncertainties
involved in extrapolating the aforementioned measurements of
the order-disorder tension from the lattice spacings they were obtained at
to the continuum limit. Current efforts to accurately determine the dual
string tension at the deconfinement transition in $SU(3)$ lattice Yang-Mills
theory \cite{vetto} actually indicate significantly higher values
of the tension at coarse lattice spacings; however, also in this case, no
conclusive statement concerning the extrapolation to the continuum limit
can be made on the basis of the preliminary data presently available. These
data display a substantial downward trend for the dual string tension as the
lattice spacing is reduced.

In view of the current status, it seems premature to draw a final conclusion
as to whether the result obtained here within the random vortex world-surface
model quantitatively reflects the behavior of full $SU(3)$ Yang-Mills
theory or not. On the other hand, it should be noted that it is possible to
significantly enhance the first-order character of the deconfinement
phase transition in the vortex model by introducing an additional term
into the action (\ref{action}) which favors vortex branchings; the
latter are presumably instrumental in establishing first-order behavior,
since they represent the feature which qualitatively distinguishes $SU(3)$
vortex configurations from $SU(2)$ vortex configurations. Introducing such a
mechanism has indeed proven to be efficacious in preliminary investigations
of the $SU(4)$ random vortex world-surface model, which will be reported
in a separate publication. Whether it will ultimately be necessary to
invoke this type of mechanism in the $SU(3)$ case is as yet unclear in view
of the current status of the data discussed above; however, in the $SU(4)$
case, such an additional term in the action will certainly play a role.

\section*{Acknowledgments}
The authors thank P.~de~Forcrand for fruitful correspondence and also
gratefully acknowledge the use of clusters managed by the High Performance
Computing (HPC) group of Jefferson Lab for a part of the computation.

%%%%%%%%%%%%%%%%%%%%%%%%%%%%%%%%%%%%%%%%%%%%%%%%%%%%%%%%%%%%%%%%%%%%%%%%%%%%%%%


\begin{thebibliography}{99}
\bibitem{R3} G.~'t~Hooft, Nucl.~Phys.~\textbf{B138} (1978) 1.
\bibitem{R7} H.~Reinhardt, Phys.~Lett.~\textbf{B557} (2003) 317.
\bibitem{kortplb} C.~Korthals-Altes, A.~Kovner and M.~A.~Stephanov,
Phys. Lett. {\bf B469} (1999) 205.
\bibitem{kortsym} C.~Korthals-Altes and A.~Kovner,
Phys. Rev. {\bf D 62} (2000) 096008.
\bibitem{mack} G.~Mack and V.~B.~Petkova, Ann. Phys. (NY) {\bf 123}
(1979) 442; \\
G.~Mack and V.~B.~Petkova, Ann. Phys. (NY) {\bf 125}
(1980) 117.
\bibitem{uwg} A.~Ukawa, P.~Windey and A.~H.~Guth, Phys.~Rev.~\textbf{D21}
(1980) 1013.
\bibitem{tomold} E.~T.~Tomboulis, Phys. Rev. {\bf D 23} (1981) 2371.
\bibitem{debbio1} L.~Del~Debbio, A.~Di~Giacomo and B.~Lucini,
Nucl. Phys. {\bf B594} (2001) 287.
\bibitem{tomineq} T.~G.~Kovacs and E.~T.~Tomboulis,
Phys. Rev. {\bf D 65} (2002) 074501.
\bibitem{thoo2} G.~'t~Hooft, Nucl.~Phys.~\textbf{B153} (1979) 141.
\bibitem{R1} M.~Engelhardt and H.~Reinhardt, Nucl.~Phys.~\textbf{B585}
(2000) 591.
\bibitem{R2} M.~Engelhardt, M.~Quandt, H.~Reinhardt, Nucl.~Phys.~\textbf{B685}
(2004) 227.
\bibitem{mpi} G.~Mack and E.~Pietarinen, Phys. Lett. {\bf B94}
(1980) 397; \\
G.~Mack and E.~Pietarinen, Nucl. Phys. {\bf B205} [FS5] (1982) 141.
\bibitem{kaj1} K.~Kajantie, L.~K\"arkk\"ainen and K.~Rummukainen,
Nucl. Phys. {\bf B357} (1991) 693.
\bibitem{tomb} T.~G.~Kovacs and E.~T.~Tomboulis,
Nucl. Phys. Proc. Suppl. {\bf 83} (2000) 553; \\
T.~G.~Kovacs and E.~T.~Tomboulis, Phys. Rev. Lett. {\bf 85} (2000) 704.
\bibitem{hart} A.~Hart, B.~Lucini, Z.~Schram and M.~Teper,
JHEP {\bf 0006} (2000) 040.
\bibitem{forcpep} P.~de~Forcrand, M.~D'Elia and M.~Pepe,
Phys. Rev. Lett. {\bf 86} (2001) 1438; \\
P.~de~Forcrand, M.~D'Elia and M.~Pepe,
Nucl. Phys. Proc. Suppl. {\bf 94} (2001) 494.
\bibitem{lor1} P.~de~Forcrand and L.~von~Smekal, Phys. Rev. \textbf{D66}
(2002) 011504; \\
P.~de~Forcrand and L.~von~Smekal,
Nucl. Phys. Proc. Suppl. {\bf 106} (2002) 619.
\bibitem{lor2} L.~von~Smekal and P.~de~Forcrand,
Nucl. Phys. Proc. Suppl. {\bf 119} (2003) 655.
\bibitem{forcjahn} P.~de~Forcrand and O.~Jahn,
Nucl. Phys. Proc. Suppl. {\bf 119} (2003) 649.
\bibitem{vetto} P.~de~Forcrand, B.~Lucini and M.~Vettorazzo, hep-lat/0409148.
\bibitem{rebbi} C.~Hoelbling, C.~Rebbi and V.~A.~Rubakov,
Nucl. Phys. Proc. Suppl. {\bf 73} (1999) 527; \\
C.~Hoelbling, C.~Rebbi and V.~A.~Rubakov,
Nucl. Phys. Proc. Suppl. {\bf 83} (2000) 485; \\
C.~Hoelbling, C.~Rebbi and V.~A.~Rubakov,
Phys. Rev. {\bf D 63} (2001) 034506.
\bibitem{debbio2} L.~Del~Debbio, A.~Di~Giacomo and B.~Lucini,
Phys. Lett. {\bf B500} (2001) 326.
\bibitem{hasen} A.~Hasenfratz, P.~Hasenfratz and F.~Niedermayer,
Nucl. Phys. {\bf B329} (1990) 739.
\bibitem{gonmart} A.~Gonz\'{a}lez-Arroyo and P.~Mart\'{\i}nez,
Nucl. Phys. {\bf B459} (1996) 337.
\bibitem{huang2} R.~Brower, S.~Huang, J.~Potvin, C. Rebbi and J.~Ross,
Phys. Rev. {\bf D 46} (1992) 4736.
\bibitem{gross2} B.~Grossmann, M.~L.~Laursen, T.~Trappenberg and U.-J.~Wiese,
Nucl. Phys. {\bf B396} (1993) 584.
\bibitem{kaj2} K.~Kajantie, L.~K\"arkk\"ainen and K.~Rummukainen,
Nucl. Phys. {\bf B333} (1990) 100.
\bibitem{huang1} R.~Brower, S.~Huang, J.~Potvin and C. Rebbi,
Phys. Rev. {\bf D 46} (1992) 2703.
\bibitem{gross1} B.~Grossmann, M.~L.~Laursen, T.~Trappenberg and U.-J.~Wiese,
Nucl. Phys. Proc. Suppl. {\bf 30} (1993) 869; \\
B.~Grossmann and M.~L.~Laursen,
Nucl. Phys. {\bf B408} (1993) 637.
\bibitem{iwasaki} Y.~Iwasaki, K.~Kanaya, L.~K\"arkk\"ainen, K.~Rummukainen and
T.~Yoshie, Phys. Rev. {\bf D 49} (1994) 3540.
\bibitem{aoki} Y.~Aoki and K.~Kanaya, Phys. Rev. {\bf D 50} (1994) 6921.
\bibitem{beinlich} B.~Beinlich, F.~Karsch and A.~Peikert,
Phys. Lett. {\bf B390} (1997) 268.
\bibitem{papa} A.~Papa, Phys. Lett. {\bf B420} (1998) 91.
\bibitem{preptop} M.~Engelhardt, Nucl. Phys. {\bf B585} (2000) 614; \\
M.~Engelhardt, Nucl. Phys. {\bf B638} (2002) 81.
\bibitem{bary} M.~Engelhardt, Phys. Rev. {\bf D 70} (2004) 074004.
\bibitem{Rx} M.~Engelhardt and H.~Reinhardt, Nucl.~Phys.~\textbf{B567}
(2000) 249.
\end{thebibliography}
\end{document}